\documentstyle[11pt,amsfonts,epsfig]{article}
\def \Ln{\mathop{\rm Ln}\nolimits}

\newcommand{\beq}{\begin{eqnarray}}
\newcommand{\eeq}{\end{eqnarray}}
\newcommand{\eqn}{\begin{equation}}
\newcommand{\een}{\end{equation}}
\begin{document}
\title{On the Jacobi-Metric Stability Criterion}
\author{M.A. Gonz\'alez Le\'on and J.L. Hern\'andez Pastora\\ {\small \sl Departamento de
Matem\'atica Aplicada. Universidad de Salamanca, SPAIN.}}

\date{}
\maketitle

\begin{abstract}

We investigate the exact relation existing between the stability
equation for the solutions of a mechanical system and the geodesic
deviation equation of the associated geodesic problem in the Jacobi
metric constructed via the Maupertuis-Jacobi Principle. We conclude
that the dynamical and geometrical approaches to the
stability/instability problem are not equivalent.

\end{abstract}


\section{Introduction}


In recent years, several authors \cite{Szyd1}, \cite{Szyd2},
\cite{Szyd3}, \cite{Pettini}, have formulated geometrical criteria
of (local) stability/instability in mechanical systems using
different \lq\lq geometrization" techniques (Maupertuis-Jacobi
Principle, Eisenhart metric, etc). The main idea is to interpret the
local instability problem, understood in terms of sensitive
dependence on initial conditions, as the study of an appropriate
geodesic deviation equation. As a principal application, chaotic
behaviors in Hamiltonian mechanical systems that appears in
cosmological models have been described using these results. Most of
these works are constructed using the Maupertuis-Jacobi principle
for natural mechanical systems, both in the very well known
Riemannian case, but also in the recent generalization to the
non-Riemanian one \cite{Szyd2}.

The Maupertuis-Jacobi principle establishes, in his classical
formulation, the equivalence between the resolution of the
Euler-Lagrange equations of a natural Hamiltonian dynamical system
(hence the Newton equations), for a given value of the mechanical
energy, and the calculation of the geodesic curves in an associated
Riemannian manifold. Throughout the time this equivalence has been
used for different purposes, as the mentioned description of chaotic
situations, but also in the analysis of ergodic systems
 \cite{Szyd3}, \cite{Anosov}, non-integrability problems \cite{Kozlov},
determination of stability properties of solitons
\cite{nosotrosNonli}, \cite{oviedo}, etcetera.

The linealization of the geodesic equations in a given manifold
gives in a natural way the so-called Jacobi equation, or geodesic
deviation equation, that allows to compute the stability/instability
of a given geodesic curve in terms of the sign of the curvature
tensor over the geodesic (in fact, for two-dimensional manifolds,
the problem reduces simply to the computation of the sign of the
gaussian curvature along the geodesic, see for instance
\cite{Laugwitz}).

The geometrization of the mechanical problem provides, as mentioned,
a possible criterion of stability of the solutions of in terms of
the geodesic deviation equation of the Jacobi metric associated to
the system, via the Maupertuis-Jacobi principle, that we will call
Jacobi-metric stability criterion.

From the point of view of the Variational Calculus applied to
geodesics, a similar result is obtained for the problem of
calculation of fixed-endpoints geodesics, where the sign of the
second variation functional is determined by the geodesic deviation
operator.


In this work we analyze the exact relation existing between this
Jacobi-metric criterion and the direct analysis of the stability of
the solutions without using the geometrization principle. The
linealization of the Euler-Lagrange equation (in this case, Newton
equations) lead to a Jacobi-like equation that generalizes the
geodesic deviation one to the case of natural mechanical systems. In
fact, this equation is also called Jacobi equation in the context of
second-order ordinary differential equations theory or KCC theory
(Kosambi-Cartan-Chern), \cite{KCC}, \cite{Antonelli}.

As we will see, the two approaches (geometrical and dynamical) are
not equivalent in general, and the Jacobi-metric criterion do not
provide exactly the same result as the standard (or dynamical) one.

The structure of the paper is as follows: in section 2 we present
the concepts involved in the work; Section 3 is dedicated to
Jacobi-metric stability criterion and its relation with the
dynamical one. In Section 4, the analysis is extended to the
variational point of view for fixed end-points problems. Finally, an
Appendix is included with several technical formulas (more or less
well known) about the behavior of covariant derivatives and
curvature tensor under conformal transformations and
reparametrizations of curves.

\section{Preliminaries and Notation}

We treat in this work with natural Hamiltonian dynamical systems,
i.e., the triple $(M,g,{\cal L})$, where $(M,g)$ is a Riemannian
manifold, and ${\cal L}$ is a natural Lagrangian function: ${\cal
L}: TM\to {\Bbb R}$, ${\cal L}=T-U$,
\[
T=\frac{1}{2} \left\langle \dot{\gamma},\dot{\gamma}\right\rangle
=\frac{1}{2}\,  g_{ij} \dot{q}^i\dot{q}^j
\]
in a system of local coordinates $(q^1,\dots,q^n)$ in $M$, $U$ is a
given smooth function $U:M\to {\Bbb R}$, $\gamma(t)\equiv
(q^1(t),\dots,q^n(t))$ is a smooth curve on $M$, and $g_{ij}$ are
the components of the metric $g$ in this coordinate system (Einstein
convention about sum in repeated indices will be used along the
paper).

The solutions (trajectories) of the system are the extremals of the
action functional $S[\gamma]$, defined in the space of smooth curves
on $M$: $\gamma: [t_0,t_1]\to M$, (we assume that $\gamma$ is at
least of class $C^2$ in the interval $(t_0,t_1)$).
\begin{equation}
S[\gamma]=\int_{t_0}^{t_1} {\cal L}( \gamma, \dot{\gamma})\,
dt\label{action}
\end{equation}
where $\dot{\gamma}\in \Gamma(TM)$ stands for the tangent
vectorfield $\frac{d\gamma}{dt}$, i.e. $\dot{\gamma}(t)\equiv
\frac{d\gamma}{dt}(t)\in T_{\gamma(t)}M$.

Euler-Lagrange equations associated to this functional are Newton
equations for the system:
\begin{equation}
\delta S=0\Rightarrow \nabla_{\dot{\gamma}}\dot{\gamma}=-{\rm grad}
U\label{newton}
\end{equation}
where $\nabla_{\dot{\gamma}}$ stands for the covariant derivative
along $\gamma(t)\equiv (q^i(t))$:
\[
\nabla_{\dot{\gamma}}\dot{\gamma}\equiv \left(
\frac{D\dot{q}^i}{dt}\right)= \left(
\frac{d\dot{q}^i}{dt}+\Gamma_{jk}^i \dot{q}^j \dot{q}^k\right)
\]
being $\Gamma_{jk}^i$ the Christofell symbols of the Levi-Civitta
connection associated to the metric $g$.
\[
\Gamma_{ij}^l=\frac{1}{2} g^{kl} \left( \frac{\partial
g_{jk}}{\partial q^i}+\frac{\partial g_{ik}}{\partial
q^j}-\frac{\partial g_{ij}}{\partial q^k}\right)
\]
grad$U$ is the vectorfield with components: $\left({\rm grad}
U\right)^i=g^{ij} \frac{\partial U}{\partial q^j}$. Equation
(\ref{newton}) is thus written in local coordinates as the following
system of ordinary differential equations:
\begin{equation}
\frac{D\dot{q}^i}{dt}=\ddot{q}^i+\Gamma_{jk}^i
\dot{q}^j\dot{q}^k=-g^{ij} \frac{\partial U}{\partial
q^j}\label{newtoncomp}
\end{equation}

Natural Hamiltonian dynamical systems over Riemannian manifolds
satisfy Legendre's condition in an obvious way, and thus the
Legendre transformation is regular, i.e. there exists a
diffeomorphism between the tangent and cotangent bundles of $M$ in
such a way that the Euler-Lagrange equations are equivalent to the
Hamilton (or canonical) equations.
\begin{equation}
\dot{p}_i=-\frac{\partial H}{\partial q^i}, \qquad
\dot{q}^j=\frac{\partial H}{\partial p_j}\label{Hamilton}
\end{equation}
\noindent where
\[
p_j=\frac{\partial {\cal L}}{\partial \dot{q}^j}=g_{ij}
\dot{q}^i;\qquad H=\frac{1}{2}  g^{ij}p_ip_j+U
\]
and $g^{ij}$ denotes the components of the inverse of $g$.

This kind of systems are autonomous, thus the mechanical energy is a
first integral of the system:
\[
E=\frac{1}{2} g_{ij}\dot{q}^i \dot{q}^j+U(q^1,\dots,q^n)
\]

Stability of the solutions of (\ref{newtoncomp}), understood in
terms of sensitive dependence on initial conditions, is interpreted
as follows: The trajectory $\gamma(t)$, solution of
(\ref{newtoncomp}),
is said to be stable if all trajectories with sufficiently close
initial conditions at $t_0$ remains close to the trajectory
$\gamma(t)$ for later times $t>t_0$.

Let $\gamma(t;\alpha)=(q^1(t;\alpha), \dots, q^n(t;\alpha))$ be a
family of solutions of  equations (\ref{newtoncomp}), with
$\gamma(t)\equiv \gamma(t;0)$, and given initial conditions
$q^i(t_0;\alpha)$, $\dot{q}^i(t_0;\alpha)$. Let us assume that the
initial conditions are analytic in the parameter $\alpha$. Then:
{\sl $\gamma(t)=(q^i(t))$ is a stable trajectory if for any
$\varepsilon>0$, there exists a $\delta(\varepsilon)>0$ such that
$|q^i(t;\alpha)-q^i(t)|<\varepsilon$ for $t>t_0$ and for all
trajectories $q(t;\alpha)=(q^i(t;\alpha))$ satisfying both
$|q^i(t_0;\alpha)-q^i(t_0)|<\delta$ and
$|\dot{q}^i(t_0;\alpha)-\dot{q}^i(t_0)|<\delta$.}

Assuming that $g$ is smooth and considering that $\gamma(t;\alpha)$
are analytic in $\alpha$ (they are solutions of an analytic system
of differential equations), we can write, for $\alpha$ sufficiently
small:
\begin{equation}
q^i(t;\alpha)= q^i(\alpha)+ \alpha\, v^i(t)+o(\alpha^2)\qquad
,\qquad v^i(t)=\left.\frac{\partial q^i(t;\alpha)}{\partial
\alpha}\right|_{\alpha=0}
\end{equation}

In a similar way, we can write:
\begin{eqnarray}
\Gamma_{jk}^i(q(t;\alpha))&=&\Gamma_{jk}^i(q(t))+ \alpha \,
\frac{\partial \Gamma_{jk}^i}{\partial q^l}(q(t))\,
v^l(t)+o(\alpha^2)\label{alphaG}\\
g^{ij}(q(t;\alpha))&=&g^{ij}(q(t)) +\alpha \,  \frac{\partial
g^{ij}}{\partial q^l}(q(t)) \, v^l(t)+o(\alpha^2)\label{alphag}
\\
\partial_j U(q(t;\alpha))&=&\partial_j U(q(t))+\alpha\, \partial_l\partial_j
U(q(t))\, v^l(t)+o(\alpha^2)\label{alphaU}
\end{eqnarray}
where $\partial_jU=\frac{\partial U}{\partial q^j}$.

Thus equations (\ref{newtoncomp}) become:
\begin{equation}
\ddot{v}^i+2\Gamma_{jk}^i \dot{v}^j \dot{q}^k=-g^{ip} v^l \left(
\partial_l\partial_p U+\Gamma_{lp}^j \partial_j U\right) +g^{jp}
\Gamma_{lp}^i\partial_j U v^l
\end{equation}
where all functions are taken at $\gamma(t)$. Taking into account
the expression of the second order covariant derivatives:
\[
\frac{D^2v^i}{dt^2}=\ddot{v}^i+\partial_l \Gamma_{jk}^i
\dot{q}^l\dot{q}^j v^k+2\Gamma_{jk}^i \dot{v}^j\dot{q}^k
+\Gamma_{jk}^iv^j\ddot{q}^k+\Gamma_{lp}^i\Gamma_{jk}^l
\dot{q}^j\dot{q}^pv^k
\]
and the components of the Riemann curvature tensor: $
R(X,Y)Z=-\nabla_X(\nabla_Y Z)+\nabla_Y(\nabla_X Z)+\nabla_{[X,Y]}Z$,
$\forall X,Y,Z\in \Gamma(TM)$:
\[
R_{lkj}^i=\Gamma_{kp}^i\Gamma_{pl}^i-\Gamma_{pl}^i\Gamma_{jk}^p+\partial_k
\Gamma_{jl}^i-\partial_l\Gamma_{jk}^i
\]
we finally arrive to the expression:
\[
\frac{D^2v^i}{dt^2}+R_{ljk}^i \dot{q}^l\dot{q}^j v^k=-g^{ij} \left(
\partial_l\partial_jU-\Gamma_{jl}^r\partial_rU\right) \, v^l
\]
that can be written as a vector equation:
\begin{equation}
\nabla_{\dot{\gamma}}\nabla_{\dot{\gamma}}
V+K_{\dot{\gamma}}(V)+\nabla_V {\rm grad} U=0\label{hesseq}
\end{equation}
where $V=V(t)\equiv (v^i(t))$, and we have used the sectional
curvature tensor:
\[
K_X(Y)=R(X,Y)X, \quad \forall X,Y\in \Gamma(TM)
\]
and the Hessian of the potential energy $U$: ${\cal H}(U)=\nabla dU$
\[
\nabla dU= \left(\partial_j\partial_lU-\partial_k U\,
\Gamma_{jl}^k\right)dq^j\otimes dq^l
\]
in such a way that $\forall X,Y\in \Gamma(TM)$
\[
\nabla dU(X,Y)=\langle \nabla_X {\rm grad}(U),Y\rangle=\langle
\nabla_Y {\rm grad}(U),X\rangle
\]

Solutions of equation (\ref{hesseq}) determine the behavior of the
family of solutions $\gamma(t,\alpha)$ with respect to the selected
solution $\gamma(t)$. Thus typical solutions of linear equations
(trigonometric functions, exponentials, etc.) will prescribe the
stability/instability situations. In several contexts equation
(\ref{hesseq}) is usually called Jacobi equation, by analogy with
the geodesic case. In fact, in the so-called KCC theory on second
order differential equations, equation (\ref{hessian}) is nothing
but the Jacobi equation for the special case of Newton differential
equations. In order to avoid confusions we will denote Hessian
operator for the mechanical system to:
\[
\Delta V=\nabla_{\dot{\gamma}}\nabla_{\dot{\gamma}}
V+K_{\dot{\gamma}}(V)+\nabla_V {\rm grad} U
\]
and thus we reserve the term Jacobi operator (and equation) to the
geodesic case, i.e. to the geodesic deviation equation.

In the special case of fixed starting point for the family of
solutions $\gamma(t;\alpha)$, i.e. $\gamma(t_0;\alpha)=\gamma(t_0)$,
an equivalent approach to equation (\ref{hessian}) can be
considered. The first variational derivative of functional
(\ref{action}) lead to Euler-Lagrange equations (\ref{newtoncomp}),
and thus the second variation functional (or Hessian functional)
will determine (together obviously with the Legendre straightness
condition, automatically satisfied for this kind of systems, see
\cite{Giaquinta}) the local minimum/maximum character of a solution
of (\ref{newtoncomp}). The second-variation functional of the action
$S$, for the case of proper variations ($V\in \Gamma(TM)$ such that
$V(t_0)=V(t_1)=0$) is:
\begin{equation}
\delta^2 S[\gamma(t)]=-\int_{t_0}^{t_1} dt \, \left\langle
\nabla_{\dot{\gamma}}\nabla_{\dot{\gamma}} \,
V+K_{\dot{\gamma}}(V)+\nabla_V {\rm grad} U,V\right\rangle
=-\int_{t_0}^{t_1} dt \left\langle \Delta V,V\right\rangle
\label{hessian}
\end{equation}
and thus the positive or negative definiteness of the $\Delta$
operator determines the character of the solution $\gamma(t)$.

\section{The Jacobi-Metric Stability Criterion}

The Maupertuis-Jacobi Principle establishes the equivalence between
the resolution of the Newton equations (\ref{newtoncomp}) of the
natural system and the calculation of the geodesic curves in an
associated Riemannian manifold. The crucial point of the Principle
is the existence of the mechanical energy as first integral for
equations (\ref{newtoncomp}). Solutions of (\ref{newtoncomp})
corresponding to a fixed value $E=T+U$ will be in one to one
correspondence with the solutions of the equations of geodesics in
the manifold $M$ with the so-called Jacobi metric: $h=2(E-U) g$,
associated to the $E$ value.

Geodesics in the Riemannian manifold $M\equiv (M,h)$\footnote{We
will call $s\equiv s_h$, i.e.: $ds_g^2= g_{ij}dq^i dq^j$, $ds^2=
h_{ij}dq^i dq^j$, and $h_{ij}=2(E-U) g_{ij}$. We will also write
$\nabla^J$ for the covariant derivative with respect to $h$, and,
for any vectorfields $X,Y\in \Gamma (TM)$: $ h(X,Y)=\left\langle
X,Y\right\rangle^J$, and $ \| X\|^J=\sqrt{\left\langle
X,X\right\rangle^J}$.} can be viewed as extremals of the free-action
functional $S_0$ or of the Length functional $L$:
\begin{equation}
S_0[\gamma]=\int_{t_0}^{t_1} \frac{1}{2} (\| \dot{\gamma}(t)\|^J)^2
\, dt;\qquad L[\gamma]=\int_{t_0}^{t_1} \| \dot{\gamma}(t)\|^J \,
dt\label{uno}
\end{equation}
for any differentiable curve $\gamma:[t_0,t_1]\to M$ connecting the
points $\gamma(t_0)=P$ and $\gamma(t_1)=Q$, $P,Q\in M$. The extremal
conditions, $\delta S_0=0$ and $\delta L=0$, lead us to the
Euler-Lagrange equations (equations of the geodesics in $M$):
\begin{equation}
\delta S_0=0\Rightarrow \nabla_{\dot{\gamma}}^J\dot{\gamma}=0;\quad
\delta L=0 \Rightarrow
\nabla_{\dot{\gamma}}^J\dot{\gamma}=\lambda(t) \dot{\gamma}, \quad
\lambda(t)=-\frac{d^2t}{ds^2}\, \left(
\frac{ds}{dt}\right)^2\label{dos}
\end{equation}
$\delta L=0$ leads to the equations of the geodesics parametrized
with respect to an arbitrary parameter $t$ (often called
pre-geodesics) as a natural consequence of the invariance under
reparametrizations of the Length functional, whereas $\delta S_0=0$
produces the equations of affinely parametrized geodesics. If we
restrict to the arc-length parametrization and we will denote, as
usual, $\gamma'=\frac{d\gamma}{ds}$, equations (\ref{dos}) are
written as: $\nabla_{\gamma'}^J\gamma'=0$, or explicitly, in terms
of Christoffel symbols $\tilde{\Gamma}_{jk}^i$ of the Levi-Civitta
connection of $h$, as:
\begin{equation}
\frac{D(q^i)'}{ds}=(q^i)''+ \tilde{\Gamma}_{jk}^i
(q^j)'(q^k)'=0\label{geodesics}
\end{equation}

The Maupertuis-Jacobi Principle can be formulated in the following
form:

\noindent {\bf Theorem of Jacobi.} {\sl The extremal trajectories of
the variational problem associated to the functional {\rm
(\ref{action})} with mechanical energy $E$, are pre-geodesics of the
manifold $(M,h)$, where $h$ is the Jacobi metric: $h=2(E-U) \, g$.}

From an analytic point of view, the theorem simply establishes that
the Newton equations (\ref{newtoncomp}) for the action $S$, are
written as the geodesic equations in $(M,h)$:   $ \nabla_{\gamma'}^J
\gamma'=0$, when the conformal transformation: $h=2(E-U)\, g$, and a
reparametrization (from the dynamical time $t$ to the arc-length
parameter $s$ in $(M,h)$) are performed.

Moreover, the dependence between the two parameters is determined
over the solutions by the equation:
\begin{equation}
\frac{ds}{dt}=2\sqrt{E-U({\gamma}(s))
T}=2(E-U(\gamma(s)))\label{eqparameter}
\end{equation}

The proof of this theorem can be viewed in several references (see
for instance \cite{Laugwitz}, see also \cite{Giaquinta}
for a general version of the Principle). However, a very
simple proof of the theorem can be carried out by the explicit
calculation of equations (\ref{geodesics}) in terms of the original
metric $g$, making use of Lemmas 1 and 2 of the Appendix, that
detail the behavior of the covariant derivatives under conformal
transformations and re-parametrizations.
$\nabla_{\gamma'}^J\gamma'=0$ turns out to be
\begin{equation}
\nabla_{\gamma'}\gamma'+\langle {\rm grad}(\ln (2(E-U))),
\gamma'\rangle \gamma'-\frac{1}{2} \langle \gamma',\gamma' \rangle
{\rm grad}(\ln (2(E-U)))=0 \label{dem2}
\end{equation}
in terms of the $\nabla$ derivative. By  applying now Lemma 2 to
(\ref{dem2}) we obtain, after the corresponding reparametrization
and simplifications, the equation
\[
\nabla_{\dot{\gamma}}\dot{\gamma}+{\rm grad} (U)=0
\]
i.e. the Newton equations of the mechanical system.

This result allows to define the Jacobi-metric criterion for
stability of the mechanical solutions in terms of the corresponding
geodesics of the Jacobi metric.

In an analogous way to the previous section, one can linearize the
equations (\ref{geodesics}) of the geodesics in $(M,h)$ by
considering a family of geodesics $\gamma(s;\alpha)$:
\[
\gamma(t;\alpha)= \gamma(t)+\alpha \, V+o(\alpha^2)
\]
with $V(s)=\left.\frac{\partial
\gamma(s;\alpha)}{\partial\alpha}\right|_{\alpha=0}$. Following the
same steps, one finally arrives to the expression
\begin{equation}
\nabla^J_{\gamma'}\nabla^J_{\gamma'}
V+K^J_{\gamma'}(V)=0\label{Jaceq}
\end{equation}
where $V=V(s)\equiv (v^i(s))$, and $K^J$ is the sectional curvature
tensor of the $h$ metric.

Equation (\ref{Jaceq}) is the Geodesic Deviation Equation, or Jacobi
Equation, for a given geodesic $\gamma(s)$ of $(M,h)$. We will
denote Jacobi Operator, or Geodesic Deviation Operator to:
\begin{equation}
\Delta^JV=\nabla^J_{\gamma'}\nabla^J_{\gamma'}
V+K^J_{\gamma'}(V)\label{Jacop}
\end{equation}

Thus stability of a solution of Newton equations $\gamma(t)$ will be
determined, in this criterion, if the corresponding geodesic
$\gamma(s)$ is stable, that finally leads to equation (\ref{Jaceq}).


In order to determine the exact relation existing between the
Jacobi-metric criterion and the dynamical o standard one, we will
analyze now equation (\ref{Jaceq}), by using the results about
conformal transformations and re-parametrizations included in the
Appendix.

Applying Lemma 1 and Lemma 3 (see Appendix) to the Jacobi operator
(\ref{Jacop}) and simplifying the expressions, equation
(\ref{Jaceq}) is written as:
\begin{eqnarray}
\Delta^JV&=&\nabla_{\gamma'}\nabla_{\gamma'}V+K_{\gamma'}(V)+\frac{1}{2}
\left\langle F,V\right\rangle \, \nabla_{\gamma'}\gamma'+
\left\langle F,\gamma'\right\rangle\, \nabla_{\gamma'}V-\frac{1}{2}
\left\langle \gamma',\gamma'\right\rangle\, \nabla_VF+\nonumber\\
&&+ \left( \left\langle F,\nabla_{\gamma'}V\right\rangle+\frac{1}{2}
\left\langle F,V\right\rangle\left\langle
F,\gamma'\right\rangle+\left\langle
\nabla_VF,\gamma'\right\rangle\right) \gamma'+\nonumber\\ &&+ \left(
\frac{1}{2} \left\langle
F,\nabla_{\gamma'}\gamma'\right\rangle+\frac{1}{2} \left\langle
F,\gamma' \right\rangle^2-\frac{1}{4} \left\langle \gamma',\gamma'
\right\rangle \left\langle F,F \right\rangle
\right) V+\nonumber\\
&&+ \left( -\frac{1}{2} \left\langle \nabla_{\gamma'}\gamma',V
\right\rangle-\left\langle
\gamma',\nabla_{\gamma'}V\right\rangle-\frac{1}{2} \left\langle
F,\gamma' \right\rangle\left\langle \gamma',V \right\rangle  \right)
F \label{paso1}
\end{eqnarray}
depending only on the metric $g$, and where $F$ denotes: $F={\rm
grad} \ln (2(E-U))$. Re-parametrization of $\gamma(s)$ in terms of
the $t$-parameter:
\[
\gamma'(s)=\frac{1}{2(E-U(\gamma(t)))}\dot{\gamma}(t),\quad
\nabla_{\gamma'}X=\frac{1}{2(E-U(\gamma(t)))} \nabla_{\dot{\gamma}}X
\]
and application of Lemma 2 to (\ref{paso1}) lead to:
\begin{eqnarray}
\Delta^JV&=&\frac{1}{(2(E-U))^2} \left(
\nabla_{\dot{\gamma}}\nabla_{\dot{\gamma}}V+K_{\dot{\gamma}}(V)+\frac{1}{2}
\left\langle F,V\right\rangle
\nabla_{\dot{\gamma}}\dot{\gamma}-\frac{1}{2} \left\langle
\dot{\gamma},\dot{\gamma}\right\rangle \nabla_VF+\right.\nonumber\\
&&+\left( \left\langle F,\nabla_{\dot{\gamma}}
V\right\rangle+\left\langle
\nabla_VF,\dot{\gamma}\right\rangle\right) \, \dot{\gamma}+\nonumber\\
&&+\left( \frac{1}{2} \left\langle F,\nabla_{\dot{\gamma}}
\dot{\gamma}\right\rangle-\frac{1}{4} \left\langle
\dot{\gamma},\dot{\gamma}\right\rangle \left\langle
F,F\right\rangle\right) V+\nonumber\\ &&\left. +\left( -\frac{1}{2}
\left\langle V,\nabla_{\dot{\gamma}} \dot{\gamma}\right\rangle-
\left\langle \dot{\gamma},\nabla_{\dot{\gamma}}
V\right\rangle\right) F\right)\label{paso2}
\end{eqnarray}

Expression (\ref{paso2}) is written in terms of quantities depending
only on the metric $g$ and the $t$-parameter. In order to relate
this expression with the Hessian operator $\Delta$ we need to
remember that $\gamma(t)$ is a solution of the Newton equations
(\ref{newtoncomp}) of energy $E$, and thus:
$\nabla_{\dot{\gamma}}\dot{\gamma}=-{\rm grad} U$, $\left\langle
\dot{\gamma}, \dot{\gamma}\right\rangle=2(E-U(\gamma(t)))$. Using
these facts and simplifying we arrive to:
\begin{equation}
\Delta^JV= \frac{1}{(2(E-U))^2} \, \left[ \Delta V-
\frac{d}{dt}\left( \frac{\left\langle V,{\rm grad} U\right\rangle
}{E-U} \right) \, \dot{\gamma} + \frac{\left\langle {\rm grad}
U,V\right\rangle + \left\langle \dot{\gamma},\nabla_{\dot{\gamma}}
V\right\rangle}{E-U} \,  {\rm grad} U\right]\label{paso4}
\end{equation}
where we have used the identity: $\left\langle \dot{\gamma},
\nabla_V {\rm grad} U\right\rangle = \left\langle V,
\nabla_{\dot{\gamma}} {\rm grad} U\right\rangle$.

Obviously, the two operators do not coincide, and correspondingly
solutions of the Jacobi equation $\Delta^JV=0$ and the equation
$\Delta V=0$ do not so. The two criteria of stability are not
equivalent. In order to investigate equation (\ref{paso4}) to
determine the reasons of this non-equivalence between the two
criteria, we have to remark that whereas all the geodesics
$\gamma(s;\alpha)$ considered in the calculation of $\Delta^J$
correspond to mechanical solutions of energy $E$ (they are solutions
of the equation of geodesics in $(M,h)$, with $h=2(E-U) g$), the
solutions $\gamma(t;\alpha)$ are in principle of energy:
\begin{equation}
E_\alpha=\frac{1}{2} \dot{q}^i(t;\alpha) g_{ij}(\gamma(t;\alpha))
\dot{q}^j(t;\alpha)+U(q(t;\alpha))\label{energia}
\end{equation}
But a correct comparison between two stability criteria is only well
established if the criteria act over the same objects. Thus the
comparison is only licit if one restricts the family
$\gamma(t;\alpha)$ to verify: $E_\alpha=E$. Expanding
(\ref{energia}) in $\alpha$ we find:
\begin{equation}
E_{\alpha}=E+\alpha\, \left( \left\langle
\dot{\gamma},\nabla_{\dot{\gamma}} V+\left\langle {\rm grad} U,
V\right\rangle  \right\rangle\right)+o(\alpha^2)\label{energia2}
\end{equation}

And thus the requirement $E_{\alpha}=E$ reduces to the verification
of: $\left\langle \dot{\gamma},\nabla_{\dot{\gamma}}\right\rangle
=-\left\langle {\rm grad} U,V\right\rangle$.

Thus the relation between the Jacobi operator and the hessian
operator restricted to equal-energy variations is:
\begin{equation}
\Delta^JV=\frac{1}{(2(E-U))^2} \, \left[ \Delta V-
\frac{d}{dt}\left( \frac{\left\langle V,{\rm grad} U\right\rangle
}{E-U} \right) \, \dot{\gamma}\right]\label{paso5}
\end{equation}
and the two operators are not equivalent, even considering the
equal-energy restriction.




\section{The Variational point of view}

As it has been explained in the Introduction of this work, we will
apply now the above obtained results to the special case of fixed
end-points, i.e. we will restrict our analysis to the situation
where the conditions: $\gamma(t_0)=P$ and $\gamma(t_1)=Q$, with $P$
and $Q$ fixed, are imposed . From the mechanical point of view, this
is exactly the case of the calculation of solitonic solutions in
Field Theories (see for instance \cite{nosotrosNonli}) where
asymptotic conditions determine the starting and ending points.
Using the Maupertuis-Jacobi Principle, this situation is translated
to the problem of calculating the geodesics connecting two fixed
points in the manifold $M$. We thus use the framework of the
Variational Calculus for fixed end-points problems.

The minimizing character (local minimum) of a geodesic $\gamma(s)$
connecting two fixed points is determined by the second variation
functional:
\begin{equation}
\delta^2 S_0=-\int_{s_0}^{s_1} \left\langle \Delta^J
V,V\right\rangle\, ds\, ,\quad \delta^2 L=-\int_{s_0}^{s_1}
\left\langle \Delta^J V^\perp,V^\perp\right\rangle \,
ds\label{hessact}
\end{equation}
where $\Delta^J$ is the geodesic deviation operator of $h$:
\[
\Delta^J V=\nabla_{\gamma'}^J\nabla_{\gamma'}^J \, V+ R^J(\gamma',V)
\gamma'=\nabla_{\gamma'}\nabla_{\gamma'} \, V+K_{\gamma'}^J(V)
\]
where $V\in \Gamma(TM)$ denotes any proper variation and $V^\perp$
is the orthogonal component of $V$ to the geodesic.

We will show now two theorems, in the first one it is established
the difference between the second variation functional of the
dynamical problem and the corresponding one to the free-action
functional associated to the Jacobi metric. In the second one, a
similar analysis is carried out for the Length functional.

\medskip

\noindent {\bf Theorem 1.} {\sl Let $\gamma(t)$ be an extremal of
the functional $S[\gamma]=\int_{t_0}^{t_1} \left( \frac{1}{2}
\left\langle \dot{\gamma},\dot{\gamma}\right\rangle
-U(\gamma)\right)\, dt$, and let $S_0^J[\gamma]=\int_{s_0}^{s_1}
\frac{1}{2} \left\langle \gamma',\gamma'\right\rangle^J\, ds$ be the
free-action functional of the Jacobi metric associated to
$S[\gamma]$ and corresponding to a fixed value, $E$, of the
mechanical energy, then the corresponding Hessian functionals
verify:
\begin{equation}
\delta^2 S_0^J[\gamma]=\delta^2 S[\gamma] +\int_{t_0}^{t_1} dt \, 2
\left\langle \dot{\gamma},\nabla_{\dot{\gamma}}V\right\rangle
\left\langle F,V\right\rangle\label{theorem1}
\end{equation}
where $F={\rm grad} \ln (2 (E-U))$.}

\bigskip

\noindent {\bf Theorem 2.} {\sl Let $\gamma(t)$ be an extremal of
the $S[\gamma]=\int_{t_0}^{t_1} \left( \frac{1}{2} \left\langle
\dot{\gamma},\dot{\gamma}\right\rangle -U(\gamma)\right)\, dt$
functional and let $L^J[\gamma]=\int_{s_0}^{s_1} \| \gamma' \| \, ds
$ be the length functional of the Jacobi metric associated to
$S[\gamma]$ and corresponding to a fixed value, $E$, of the
mechanical energy, then the corresponding hessian functionals
verify:
\begin{equation}
\delta^2L^J[\gamma]=\delta^2 S[\gamma]-\int_{t_0}^{t_1}
\frac{dt}{2(E-U)} \left[ \left\langle
\nabla_{\dot{\gamma}}\dot{\gamma},V\right\rangle - s \left\langle
\dot{\gamma},\nabla_{\dot{\gamma}}V\right\rangle\right]^2
\label{theorem2}
\end{equation}
}

\noindent From (\ref{theorem2}) it is obvious that minimizing
geodesics are equivalent to minimizing (stable) solutions of the
dynamical system, i.e. a positive definiteness of $\delta^2L^J$
implies the same behaviour for $\delta^2 S$, but it is not
necessarily true the reciprocal statement.

\medskip

If we restrict the variations to the orthogonal ones, $V=V^\perp$,
(\ref{theorem2}) can be re-written as:
\[
\left. \delta^2S\right|_{V=V^{\bot}}=\delta^2L^J+\int_{s_0}^{s_1} ds
\left( \langle F^J,V^{\bot}\rangle^J\right)^2
\]


The proofs of these two theorems are based on the behaviour of the
covariant derivatives and the curvature tensor under
reparametrizations and conformal transformations of the metric
tensor. We thus use the technical results included in the Appendix.

\noindent {\bf Proof of Theorem 1.}  We start with equation
(\ref{hessact}) particularized to the case of the Jacobi metric:
\[
\delta^2 S_0^J[\gamma]=\int_{s_0}^{s_1} \, ds \left\langle -\Delta^J
V,V\right\rangle^J
\]
with $\Delta^J V=\nabla_{\gamma'}^J\nabla_{\gamma'}^J
V+K_{\gamma'}^J(V)$.

Using expression (\ref{paso4}), deduced in the previous section
after changing the metric and re-parametrizing, we can write:
\begin{eqnarray}
&&\left\langle \nabla_{\gamma'}^J \nabla_{\gamma'}^J
V+K_{\gamma'}^J(V),V \right\rangle^J= \frac{1}{2 (E-U)} \left\langle
\nabla_{\dot{\gamma}}\nabla_{\dot{\gamma}}
V+K_{\dot{\gamma}}(V)+\nabla_V {\rm grad}U,V \right\rangle
+\label{pag6-2}\\ &&+\frac{1}{2(E-U)} \frac{\partial}{\partial t}
\left(  \langle F,V\rangle \langle
\dot{\gamma},V\rangle\right)+\frac{1}{(E-U)^2} \langle
\dot{\gamma},\nabla_{\dot{\gamma}}V\rangle \langle {\rm
grad}U,V\rangle \nonumber
\end{eqnarray}

\noindent And thus, the second variation functional is written as:
\begin{eqnarray*}
\frac{d^2 S_0^J[\gamma]}{d\xi^2}(0)&=& -\int_{s_0}^{s_1} ds
\left\langle
\nabla_{\gamma'}^J \nabla_{\gamma'}^J  V+K_{\gamma'}^J(V),V \right\rangle^J =\\
&& =-\int_{t_0}^{t_1} dt \left\langle
\nabla_{\dot{\gamma}}\nabla_{\dot{\gamma}}
V+K_{\dot{\gamma}}(V)+\nabla_V {\rm grad}U,V \right\rangle +\\ &&
+\int_{t_0}^{t_1} dt 2\langle
\dot{\gamma},\nabla_{\dot{\gamma}}V\rangle \langle F,V\rangle
-\left. \langle F,V\rangle \langle \dot{\gamma},V\rangle
\right|_{t_0}^{t_1}
\end{eqnarray*}

\medskip

\noindent For proper variations: $V(t_1)=V(t_2)=0$
\[
\frac{d^2S_0^J[\gamma]}{d\xi^2}(0)=\frac{d^2
S[\gamma]}{d\xi^2}(0)+\int_{t_1}^{t_2} dt\,  2 \left\langle
\dot{\gamma},\nabla_{\dot{\gamma}}V\right\rangle \left\langle
F,V\right\rangle
\]
with $F={\rm grad}\Ln (2(E-U)) = \displaystyle{ -\frac{1}{E-U} {\rm
grad} U}$.

\noindent Q.E.D.

\medskip

\noindent {\bf Proof of Theorem 2:} For the Length functional we
have:
\[
\frac{d^2L^J[\gamma]}{d\xi^2}(0)=-\int_{s_1}^{s_2} ds \left\langle
\nabla_{\gamma'}^J\nabla_{\gamma'}^J
V^{\bot}+K_{\gamma'}^J(V^{\bot}),V^{\bot}\right\rangle^J
\]
where
\[
V^{\bot}=V-\left\langle \frac{\gamma'}{\|
\gamma'\|^J},V\right\rangle^J \frac{\gamma'}{\|
\gamma'\|^J}=V-\left\langle \gamma',V\right\rangle \gamma'
\]
and thus:
\[
\frac{d^2L^J[\gamma]}{d\xi^2}(0)=\frac{d^2S_0^J[\gamma]}{d\xi^2}(0)-\int_{s_1}^{s_2}
ds \left( \left\langle
\gamma',\nabla_{\gamma'}^JV\right\rangle^J\right)^2
\]

\medskip

\noindent By using Theorem 1, we have that
\begin{eqnarray*}
\frac{d^2L^J[\gamma]}{d\xi^2}(0)&=&\frac{d^2S_0^J[\gamma]}{d\xi^2}(0)-\int_{s_1}^{s_2}
ds \left( \left\langle
\gamma',\nabla_{\gamma'}^JV\right\rangle^J\right)^2=\\ &=&
\frac{d^2S[\gamma]}{d\xi^2}(0) +\int_{t_1}^{t_2} 2 \langle
\dot{\gamma},\nabla_{\dot{\gamma}}V\rangle \langle F,V dt\rangle
+\int_{t_1}^{t_2} A(t) dt
\end{eqnarray*}
where:
\[
\int_{t_1}^{t_2} A(t) dt=-\int_{s_1}^{s_2} ds \left( \left\langle
\gamma',\nabla_{\gamma'}^JV\right\rangle^J\right)^2=-\int_{t_1}^{t_2}
dt\, (2(E-U))^3 \left\langle
\gamma',\nabla_{\gamma'}^JV\right\rangle^2
\]

\noindent From Lemma 1 and Newton equations, we have
\[
\int_{t_1}^{t_2} A(t) dt=-\int_{t_1}^{t_2} dt \frac{1}{2(E-U)}
\left( \langle \dot{\gamma},\nabla_{\dot{\gamma}}V\rangle + (E-U)
\langle F,V\rangle \right)^2
\]

\noindent Finally
\[
\frac{d^2L^J[\gamma]}{d\xi^2}(0)=\frac{d^2S[\gamma]}{d\xi^2}(0)
-\int_{t_1}^{t_2} dt \frac{1}{2(E-U)} \left[ \langle
\dot{\gamma},\nabla_{\dot{\gamma}}V\rangle-\langle
\nabla_{\dot{\gamma}}\dot{\gamma},V \rangle \right]^2
\]
Q.E.D.



\section{Appendix}

\noindent {\bf Lemma 1.} {\sl Given a conformal transformation in a
riemannian manifold, $(M,g)\to (M, \tilde{g})$; $\tilde{g}=f({\bf
x}) \, g$, $f({\bf x})\neq 0$, $\forall {\rm x}\in M$, let $\nabla$
and $\tilde{\nabla}$ be the associated Levi-Civita connections
respectively. Then, for all $X,Y,Z \in \Gamma(TM)$ it is verified
that:{\small
\begin{equation}
\tilde{\nabla}_XY=\nabla_XY+\frac{1}{2} \langle F,Y\rangle X+
\frac{1}{2} \langle F,X\rangle Y-\frac{1}{2} \langle X,Y\rangle
F\label{lema1a}
\end{equation}
\begin{eqnarray}
\tilde{\nabla}_X\tilde{\nabla}_YZ&=&\nabla_X\nabla_YZ+\frac{1}{2}\langle
F,Z \rangle \nabla_XY+\frac{1}{2} \langle F,Y\rangle
\nabla_XZ-\frac{1}{2}  \langle Y,Z\rangle\nabla_XF+\frac{1}{2}
\langle F,X\rangle \nabla_YZ+\nonumber\\ && +\left( \frac{1}{2}
\langle F,\nabla_YZ\rangle+\frac{1}{2}  \langle F,Z\rangle \langle
F,Y\rangle-\frac{1}{4}  \langle Y,Z\rangle  \langle F,F
\rangle\right) X+\nonumber\\ &&+\left( \frac{1}{2}  \langle
\nabla_XF,Z\rangle+\frac{1}{2}  \langle
F,\nabla_XZ\rangle+\frac{1}{4}  \langle F,Z\rangle  \langle
F,X\rangle\right) Y+\nonumber \\ &&+\left( \frac{1}{2} \langle
\nabla_XF,Y \rangle+\frac{1}{2}  \langle F,\nabla_XY
\rangle+\frac{1}{4} \langle F,X\rangle  \langle F,Y\rangle\right)
Z+\label{lema1b}\\ && +\left( \frac{-1}{2}  \langle
\nabla_XY,Z\rangle-\frac{1}{2} \langle
Y,\nabla_XZ\rangle-\frac{1}{2}  \langle X,\nabla_YZ\rangle
-\frac{1}{4} \langle F,Z\rangle  \langle X,Y\rangle-\frac{1}{4}
\langle F,Y\rangle  \langle X,Z\rangle\right) F\nonumber
\end{eqnarray}}
where the scalar products are taken with respect to the metric $g$
and $F={\rm grad}\left( \ln f \right)$ ({\rm grad} stands for the
gradient with respect the metric $g$).}

\medskip

\noindent {\bf Proof:}  {\small By direct calculation. Let us
consider the expression of the Christoffel symbols of the
$\tilde{g}$ metric:
\[
\tilde{\Gamma}_{jk}^i=\frac{1}{2} \tilde{g}^{ir}\left( \partial_k
\tilde{g}_{jr}-\partial_r \tilde{g}_{jk}+\partial_j \tilde{g}_{rk}
\right)
\]
and substitute $\tilde{g}_{ij}=f\, g_{ij}$,
$\tilde{g}^{ij}=\frac{1}{f} g^{ij}$. So
\[
\tilde{\Gamma}_{jk}^i=\Gamma_{jk}^i+\frac{1}{2} \left[ {\rm
grad}(\ln f)^m \left( \delta_j^i g_{mk}+\delta_k^i g_{mj}\right)
-g_{jk} {\rm grad}(\ln f)^i\right]
\]
and the covariant derivative will be
\[
\tilde{\nabla}_XY=X^j \tilde{\nabla}_jY=X^j \left( \frac{\partial
Y^i}{\partial x^j}+\tilde{\Gamma}_{jk}^i Y^k\right)
\frac{\partial}{\partial x^i}=\nabla_XY+A_{jk}^i X^jY^k
\frac{\partial}{\partial x^i}
\]
where $A_{jk}^i$ stands for:
\[
A_{jk}^iX^jY^k=\frac{1}{2} \left[ {\rm grad}(\ln f)^m \left(
\delta_j^i g_{mk}+\delta_k^i g_{mj}\right) -g_{jk} {\rm grad}(\ln
f)^i\right]X^jY^k=
\]
\[
=\frac{1}{2} \left( \langle {\rm grad}(\ln f),Y\rangle X^i+\langle
{\rm grad}(\ln f),X\rangle Y^i\right) -\frac{1}{2} \langle
X,Y\rangle {\rm grad}(\ln f)^i
\]
Finally, simplifying
\[ \tilde{\nabla}_XY=\nabla_XY+\frac{1}{2} \langle {\rm
grad}(\ln f),Y\rangle X+ \frac{1}{2} \langle {\rm grad}(\ln
f),X\rangle Y-\frac{1}{2} \langle X,Y\rangle {\rm grad}(\ln f)
\]
and similarly for (\ref{lema1b}). }

\noindent Q.E.D.

\bigskip

\noindent {\bf Lemma 2.} {\sl Given a (differentiable) curve
$\gamma:[t_1,t_2]\to M$ on $M$, let $\gamma(s)=\gamma(t(s))$ be an
admissible re-parametrization of $\gamma$, $ds=f({\bf x}(t))\, dt$
($f({\bf x}(t))\neq 0,\forall t\in[t_1,t_2]$). Then $\forall X\in
\Gamma(TM)$: {\small
\begin{equation}
\nabla_{\gamma'}X=\frac{1}{f({\bf x})
}\nabla_{\dot{\gamma}}X\label{lema2a}
\end{equation}
\begin{equation}
\nabla_{\gamma'}\gamma'=\frac{1}{f({\bf x})^2} \left(
\nabla_{\dot{\gamma}}\dot{\gamma}-\langle {\rm grad}(\ln
f),\dot{\gamma}\rangle \dot{\gamma}\right)\label{lema2b}
\end{equation}
\begin{equation}
\nabla_{\gamma'}\nabla_{\gamma'}X=\frac{1}{f({\bf x})^2} \left(
\nabla_{\dot{\gamma}}\nabla_{\dot{\gamma}} X-\left\langle {\rm
grad}(\ln f),\dot{\gamma}\right\rangle \,
\nabla_{\dot{\gamma}}X\right)\label{lema2c}
\end{equation}
} where $\dot{\gamma}(t)=\frac{d\gamma(t)}{dt}$ and
$\gamma'(s)=\frac{d\gamma(s)}{ds}$.}

\noindent {\bf Proof:} {\small Again by direct calculation
\[
\nabla_{\gamma'}X=\left( \frac{dX^i}{ds}+\Gamma_{jk}^i x'^j
X^k\right) \frac{\partial}{\partial x^i}= \left(
\frac{dX^i}{dt}\frac{dt}{ds}+\Gamma_{jk}^i \dot{x}^j\frac{dt}{ds}
X^k\right) \frac{\partial}{\partial x^i}=\frac{1}{f}
\nabla_{\dot{\gamma}}X
\]

\begin{eqnarray*}
\nabla_{\gamma'}\gamma'&&=\frac{1}{f} \nabla_{\dot{\gamma}}\gamma'=
\frac{1}{f} \left( \frac{dx'^i}{dt}+\Gamma_{jk}^i\dot{x}^j
x'^k\right) \frac{\partial}{\partial x^i} =\left(\frac{1}{f}
\frac{d}{dt} \left( \frac{\dot{x}^i}{f}\right)+\Gamma_{jk}^i
\dot{x}^j
\dot{x}^k \frac{1}{f^2} \right)\frac{\partial}{\partial x^i}=\\
&& =\frac{1}{f^2} \left( \nabla_{\dot{\gamma}}\dot{\gamma}-
\partial_k \ln f \dot{x}^k\dot{x}^i \frac{\partial}{\partial
x^i}\right)=\frac{1}{f^2}\left(
\nabla_{\dot{\gamma}}\dot{\gamma}-\langle {\rm grad}(\ln
f),\dot{\gamma}\rangle \dot{\gamma}\right)
\end{eqnarray*}

\begin{eqnarray*}
\nabla_{\gamma'}\nabla_{\gamma'}X&&= \nabla_{\gamma'} \left(
\frac{1}{f} \nabla_{\dot{\gamma}}X\right)=\nabla_{\gamma'}\left(
\frac{1}{f} \right) \nabla_{\dot{\gamma}}X+\frac{1}{f^2}
\nabla_{\dot{\gamma}}\nabla_{\dot{\gamma}} X=\\ && = \frac{dt}{ds}
\, \frac{d}{dt} \left( \frac{1}{f}\right) \nabla_{\dot{\gamma}}X+
\frac{1}{f^2} \nabla_{\dot{\gamma}}\nabla_{\dot{\gamma}}
X=\frac{1}{f^2} \left( \nabla_{\dot{\gamma}}\nabla_{\dot{\gamma}} X-
\left\langle {\rm grad}(\ln f),\dot{\gamma}\right\rangle
\nabla_{\dot{\gamma}}X\right)
\end{eqnarray*}
}

\noindent Q.E.D.

\medskip

\noindent {\bf Lemma 3.} {\sl Given a conformal transformation in a
Riemannian manifold: $(M,g)\to (M,\tilde{g})$, $\tilde{g}=f({\bf x})
g$, let $R$ and $\tilde{R}$ be the associated curvature tensors
respectively. Then, for any $X,Y,Z\in\Gamma(TM)$, it is verified
that: {\small
\begin{eqnarray} \tilde{R}(X,Y)Z&&=
R(X,Y)Z-\frac{1}{2} \langle X,Z\rangle
\nabla_YF+\frac{1}{2} \langle Y,Z\rangle \nabla_XF+\nonumber\\
&&+\left( \frac{1}{2} \left\langle \nabla_YF,Z\right\rangle
-\frac{1}{4} \left\langle F,Z\right\rangle \left\langle
F,Y\right\rangle+\frac{1}{4} \left\langle Y,Z\right\rangle
\left\langle F,F\right\rangle \right) \, X+\nonumber\\ &&+\left(
-\frac{1}{2} \left\langle \nabla_XF,Z\right\rangle +\frac{1}{4}
\left\langle F,Z\right\rangle \left\langle
F,X\right\rangle-\frac{1}{4} \left\langle X,Z\right\rangle
\left\langle F,F\right\rangle \right) \, Y+\nonumber\\ &&+\left(
\frac{1}{2} \left\langle  \nabla_YF,X\right\rangle -\frac{1}{2}
\left\langle \nabla_XF,Y\right\rangle \right) \, Z+\label{lema3}\\
&& + \left( \frac{1}{4} \left\langle F,Y\right\rangle \left\langle
X,Z\right\rangle-\frac{1}{4} \left\langle F,X\right\rangle
\left\langle Y,Z\right\rangle \right) \, F \nonumber
\end{eqnarray}
} \noindent where $\nabla$ is the Levi-Civita connection associated
to $g$, $F={\rm grad}(\ln f)$ and the scalar products and the
gradient are taken with respect to the metric $g$.}

\medskip

\noindent {\bf Proof:}  {\small Apply Lemma 1 to the formula: $
\tilde{R}(X,Y)Z=-\tilde{\nabla}_X(\tilde{\nabla}_Y
Z)+\tilde{\nabla}_Y(\tilde{\nabla}_X Z)+\tilde{\nabla}_{[X,Y]}Z$,
and simplify.

\noindent Q.E.D. }

\end{document}